\documentclass[multphys,vecphys]{svmult}

\usepackage{makeidx}         
\usepackage{graphicx}        
\usepackage{multicol}        
\usepackage[bottom]{footmisc}

\makeindex             

\begin{document}

\title*{Modeling Chandra X-ray observations of Galaxy Clusters using
  Cosmological Simulations}
\titlerunning{Modeling Chandra X-ray observations}
\author{Daisuke Nagai\inst{1}\and
Andrey V. Kravtsov\inst{2} \and Alexey Vikhlinin\inst{3,4}}
\institute{Theoretical Astrophysics, California Institute of
Technology, Mail Code 130-33, Pasadena, CA 91125 
\texttt{daisuke@caltech.edu}
\and Department of Astronomy and Astrophysics, KICP, \& EFI, The University of Chicago, 5640
South Ellis Ave., Chicago, IL 60637
\and Harvard-Smithsonian Center for Astrophysics, 60 Garden Street, 
Cambridge, MA 02138
\and Space Research Institute, 8432 Profsojuznaya St., GSP-7, Moscow
 117997, Russia }

%
%
\maketitle

\section{Abstract}
\label{sec:1}

X-ray observations of galaxy clusters potentially provide powerful
cosmological probes if systematics due to our incomplete knowledge of the
intracluster medium (ICM) physics are understood and controlled. In this
paper, we study the effects of galaxy formation on the properties of the ICM
and X-ray observable-mass relations using high-resolution self-consistent
cosmological simulations of galaxy clusters and comparing their results with
recent \emph{Chandra} X-ray observations.  We show that despite complexities
of their formation and uncertainties in their modeling, clusters of galaxies
both in observations and numerical simulations are remarkably regular outside
of their cores, which holds great promise for their use as cosmological
probes.

\section{Testing X-Ray Measurements of Galaxy Clusters}
\label{sec2}

\begin{figure}[t]
  \centering \includegraphics[height=5.8cm]{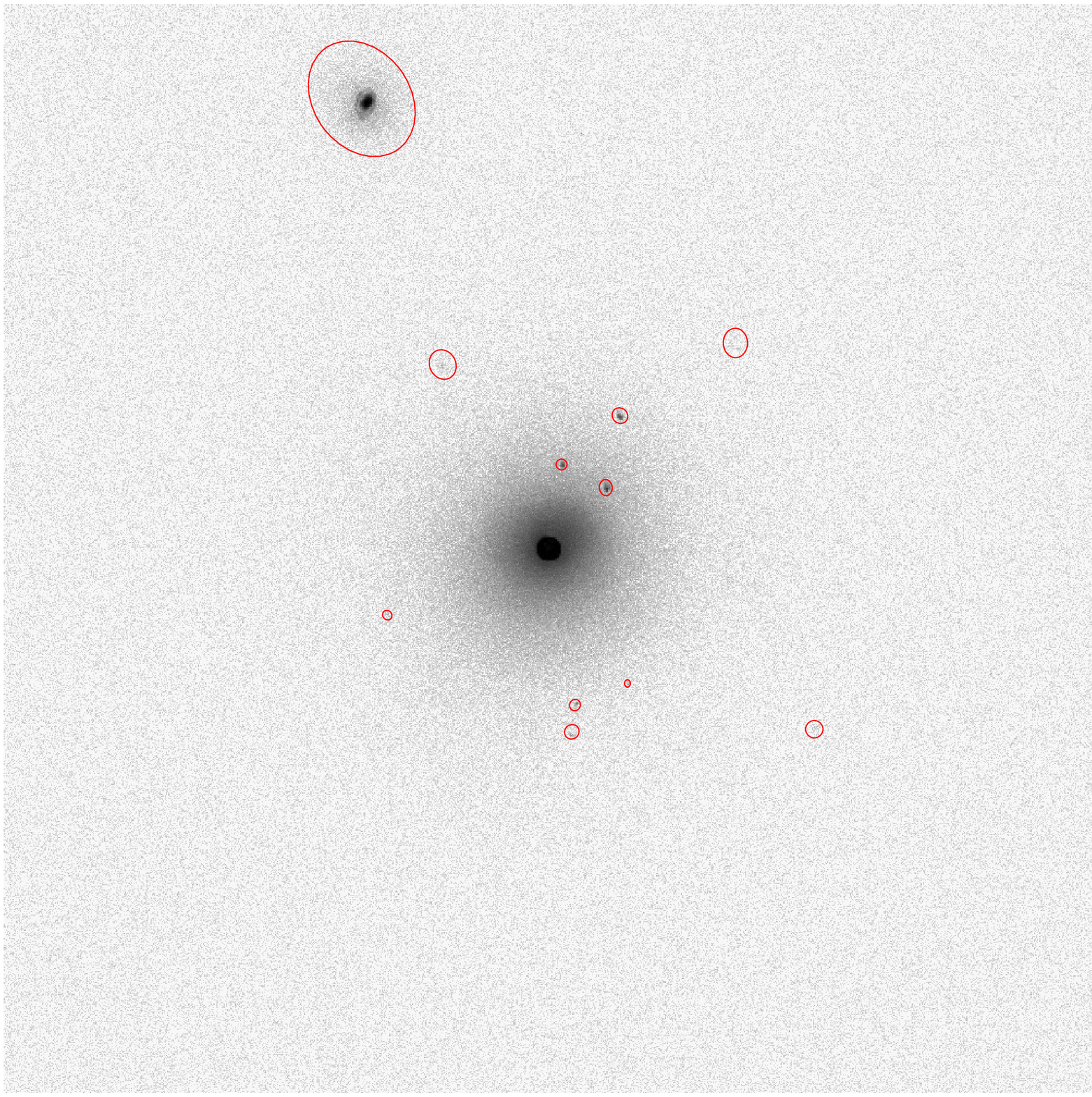}
  \includegraphics[height=5.8cm]{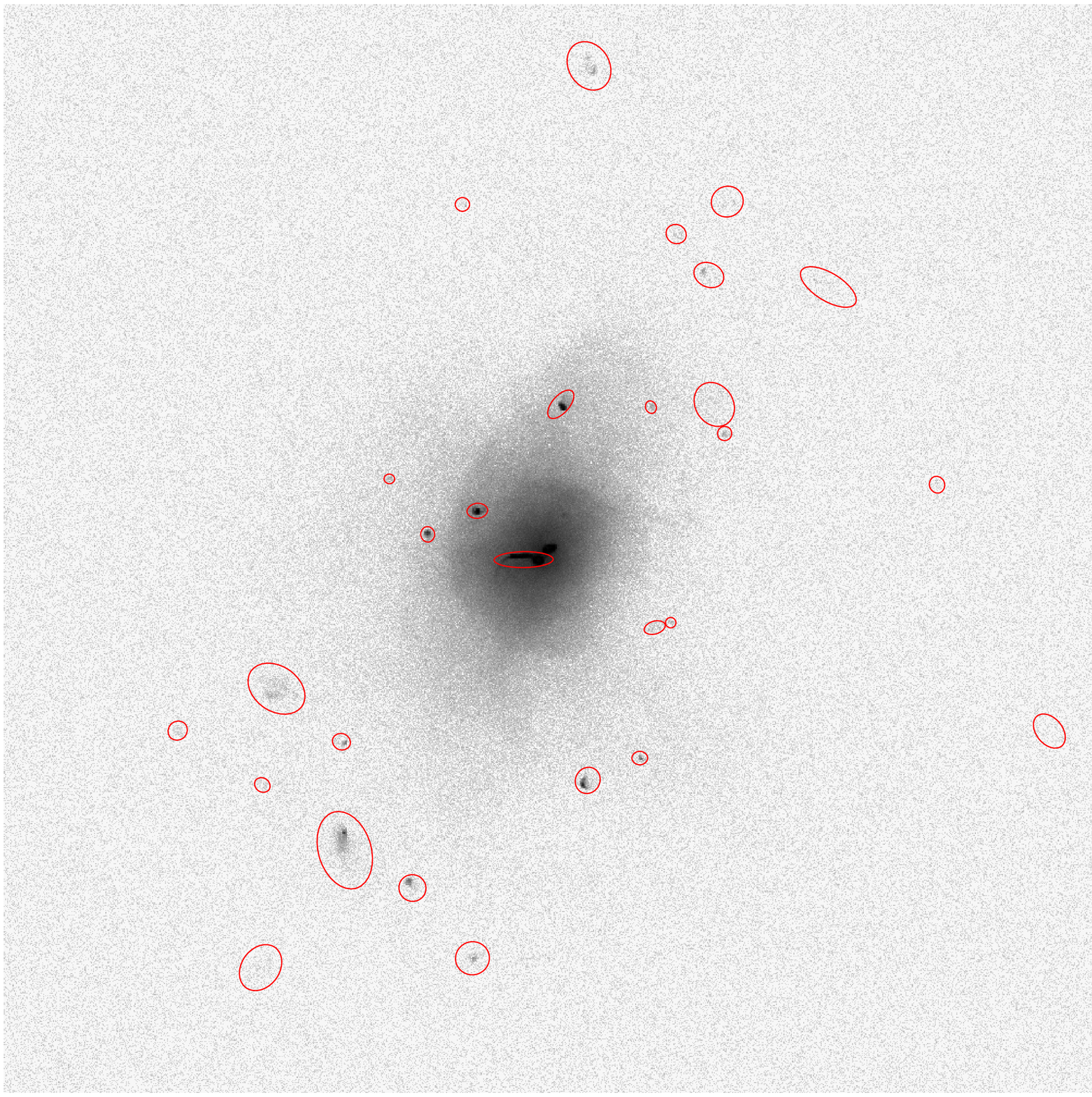}
\caption{Mock \emph{Chandra} images of one of the relaxed (\emph{left}) and
  unrelaxed (\emph{right}) simulated clusters at z=0. The detectable extended
  X-ray sources, indicated by ellipses, are detected and masked out from the
  analysis.  The physical size of the images is $5\,h^{-1}$~Mpc.}
\label{fig1}      
\end{figure}

\begin{figure}[t]
\centering
\includegraphics[height=12cm]{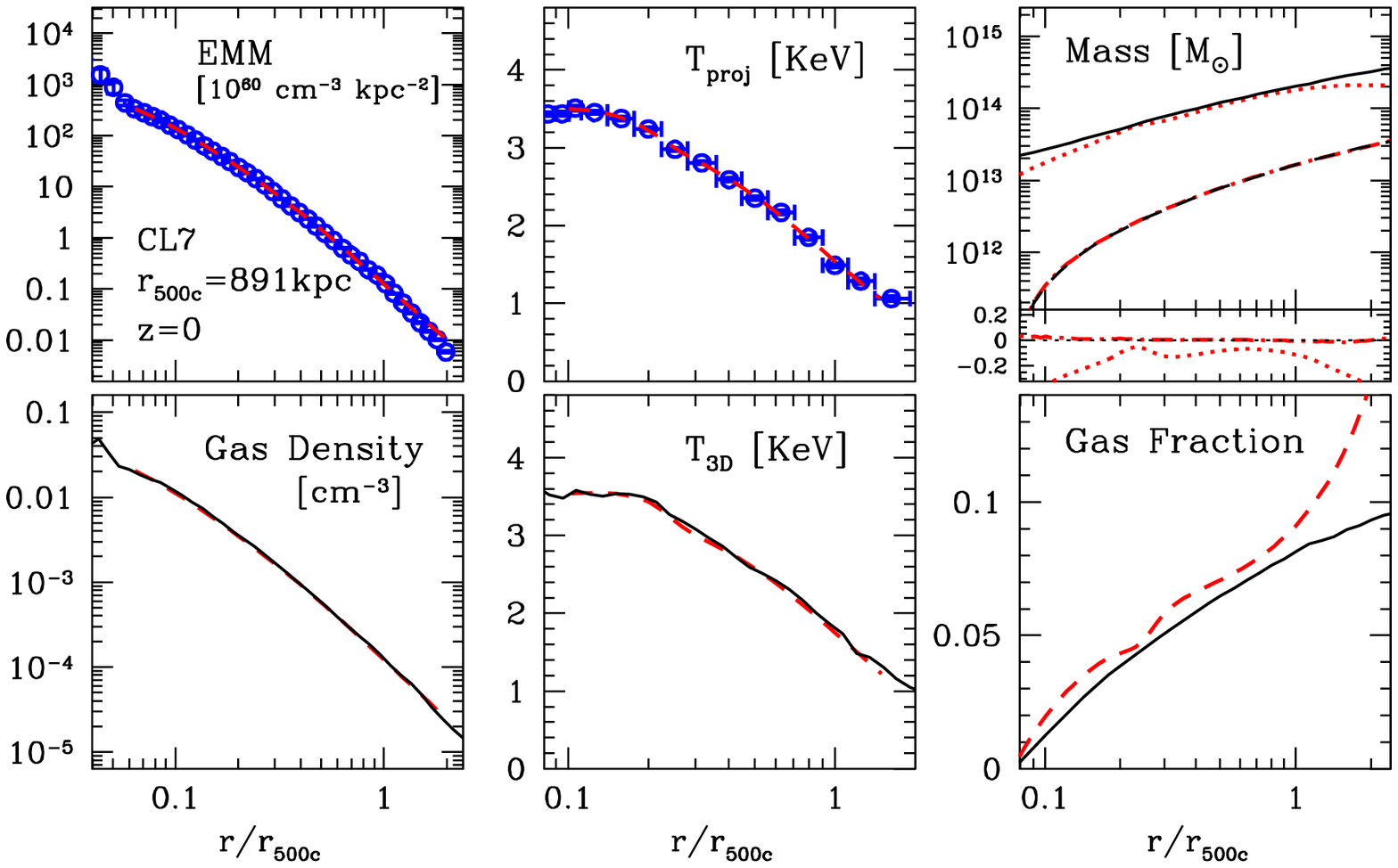}
\vspace{-5cm}
\caption{The mock \emph{Chandra} analyses of one of the relaxed clusters at $z=0$
  with $M_{\rm 500c}=1.41\times 10^{14}h^{-1}M_{\odot}$.  In the \emph{left}
  and \emph{middle} panels, the best-fit model (\emph{dashed} lines) recovers
  well both the projected profiles and the actual 3D gas profiles.  In the
  \emph{upper-right} panel, the reconstructed $M_{\rm gas}$ profile
  (\emph{dot-dashed} line) is accurate to a few percent in the entire radial
  range shown.  The hydrostatic $M_{\rm tot}$ estimate (\emph{dotted} line),
  on the other hand, is biased low by about 5\%--10\% in the radial range,
  $[0.2,1.0] r_{\rm 500c}$.  The \emph{lower-right} shows that measured
  cumulative $f_{\rm gas}$ is biased high by $\approx 10\%$ in the radial
  range of $[0.2,1.0] r_{\rm 500c}$ for this cluster, and it is primarily due
  to the bias in the hydrostatic mass estimate.}
\label{fig2}      
\end{figure}

X-ray observations with {\it Chandra} and {\it XMM-Newton } enable us to study
properties of the ICM with unprecedented detail and accuracy and provide
important handles on the ICM modeling and associated systematics. Their superb
spatial resolution and sensitivity enable accurate X-ray brightness and
temperature measurements at a large fraction of the cluster virial radius and
also make it simple to detect most of the small-scale X-ray clumps. Despite
this recent observational progress, the biases in the determination of the key
cluster properties remain relatively uncertain. 

We therefore assess the accuracy of the X-ray measurements of galaxy cluster
properties using mock \emph{Chandra} analyses of cosmological cluster
simulations and analyzing them using a model and procedure essentially
identical to that used in real data analysis
\cite{vikhlinin_etal05c,vikhlinin_etal06}. The comparison of the true and
derived cluster properties provides an assessment of biases introduced by the
X-ray analysis.  We examine the bias in mass measurements separately for
dynamically relaxed and non-relaxed clusters, identified based on the overall
structural morphology of their \emph{Chandra} images, which mimics the
procedure used by observers. The typical examples of systems classified as
relaxed or unrelaxed are shown in Figure~\ref{fig1}.  To check for any
redshift dependence in such biases, we also analyze the simulation outputs at
$z=0$ and $0.6$.  The simulations and analysis procedures are fully described
in \cite{nagai_etal06}.

Figure~\ref{fig2} illustrates that the X-ray analysis provides accurate
reconstruction of the 3D properties of the ICM for the nearby, relaxed
clusters. The strongest biases we find are those in the hydrostatic mass
estimates, which is biased at a level of about $13\%$ at $r_{500c}$ even in
the relaxed clusters.  We find that the biases are primarily due to additional
pressure support provided by subsonic bulk motions in the ICM, ubiquitous in
our simulations even in relaxed systems
\cite{faltenbacher_etal05,rasia_etal06,lau_etal06}.  These biases are related
to physics explicitly missing from the hydrostatic method (e.g., turbulence),
and not to deficiencies of the X-ray analysis.  Gas fraction determinations
are therefore biased high.  The bias increases toward cluster outskirts and
depends sensitively on its dynamical state, but we do not observe significant
trends of the bias with cluster mass or redshift. We also compute a X-ray
spectral temperature ($T_X$), a value derived from a single-temperature fit to
the integrated cluster spectrum excluding the core ($< 0.15 r_{500c}$) and
detectable small-scale clumps \cite{mazzotta_etal04,vikhlinin_etal06b}.

\section{Effects of Galaxy Formation on the ICM Profiles}
\label{sec3}

\begin{figure}[t]
  \centering \includegraphics[height=6.0cm]{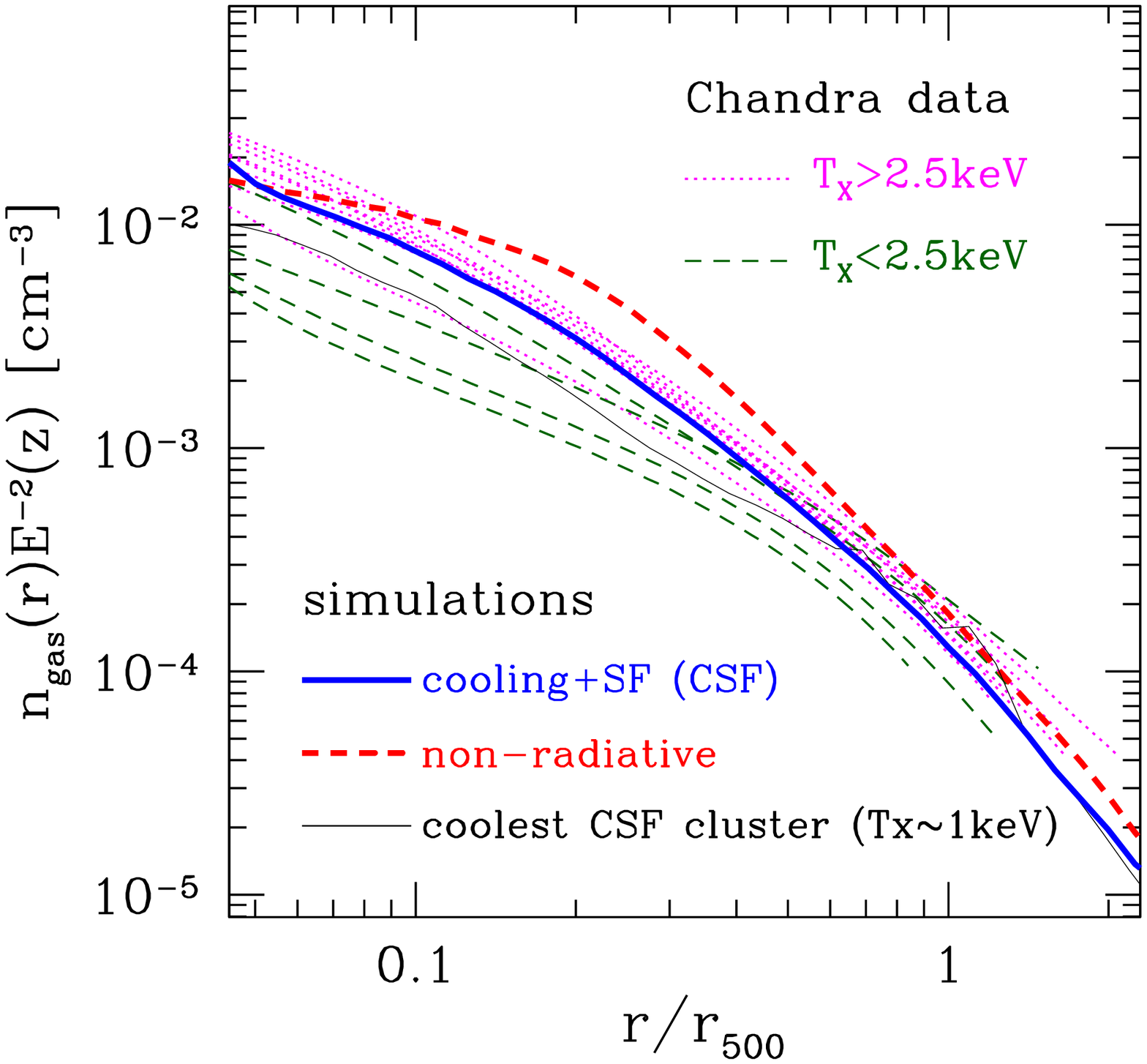} \hspace{-0.4cm}
  \includegraphics[height=6.0cm]{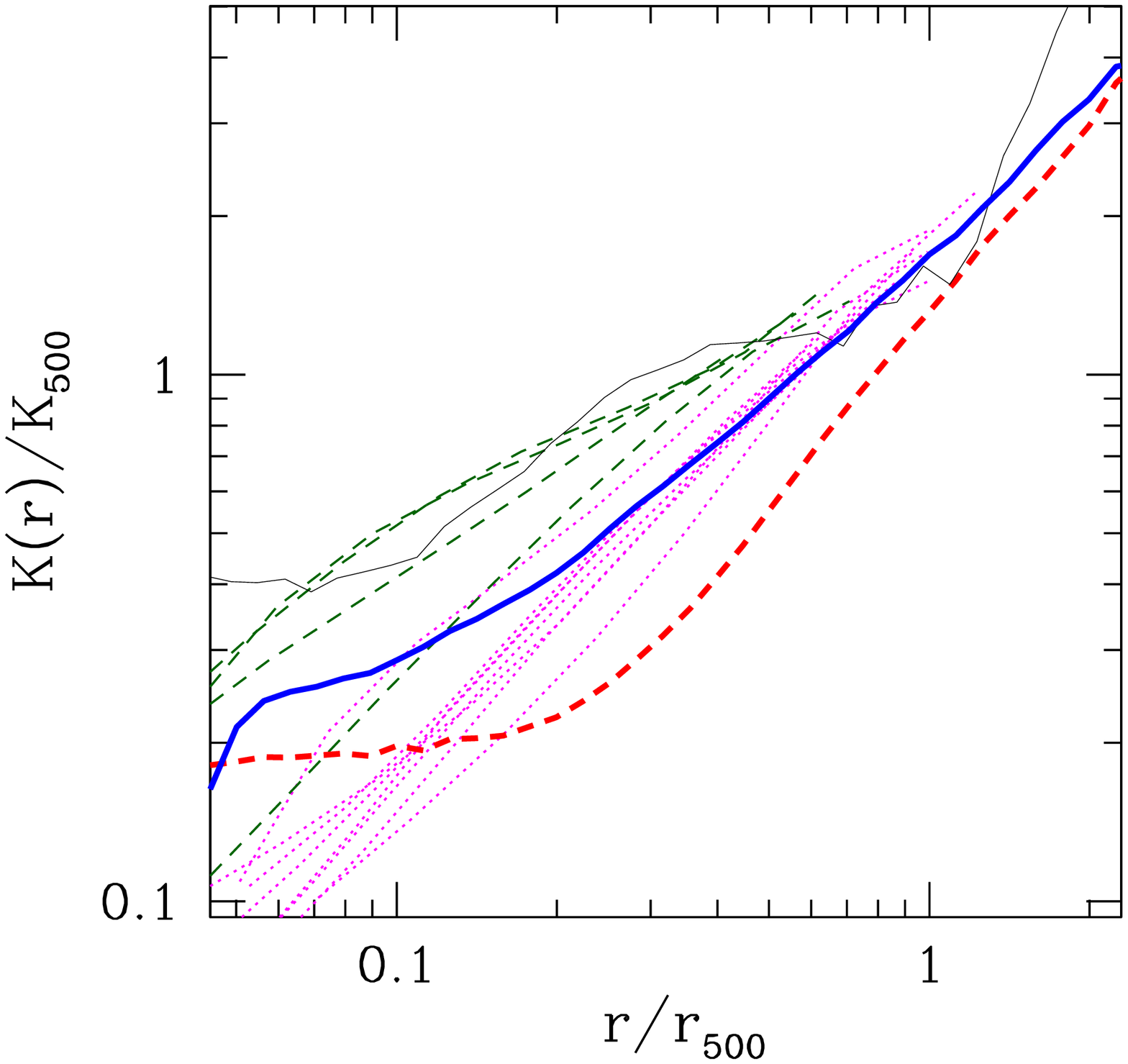}
\caption{  
  The ICM gas density and entropy profiles of simulated clusters and Chandra
  X-ray observations of nearby ($z\approx 0$) relaxed clusters.}
\label{fig3}     
\end{figure}

Next, we investigate the effects of galaxy formation on the ICM properties and
compare the results of simulations with recent \emph{Chandra} X-ray
observations of nearby relaxed clusters.  The impact of galaxy formation on
the properties of ICM are investigated by comparing simulations performed with
and without the processes of galaxy formation (e.g., gas cooling, star
formation, stellar feedback and metal enrichment), which we refer to as the
cooling+SF (CSF) and non-radiative runs, respectively.  Fig.~\ref{fig3} shows
that gas cooling and star formation modify both the normalization and the
shapes of the ICM profiles.  What happens is that the removal of low-entropy
gas in the inner region raises the level of entropy and lowers the gas density
\cite{voit_etal01}.  The effects are strongly radial dependent and increase
toward the inner regions down to about $\sim 0.1r_{500c}$, inside which the
observed properties are not well reproduced in the simulations.  On the other
hand, the ICM properties outside the cores in the cooling CSF 
simulations and observations
agree quite well.  At $r_{500c}$, both the ICM density and entropy profiles of
different mass systems converge, indicating that the clusters are self-similar
in the outskirts. Note that the non-radiative simulations predict
overall shape of the density and entropy profiles inconsistent with
observations.

\section{X-ray observable-mass relations}
\label{sec4}

\begin{figure}[t]
  \centering \includegraphics[height=6.1cm]{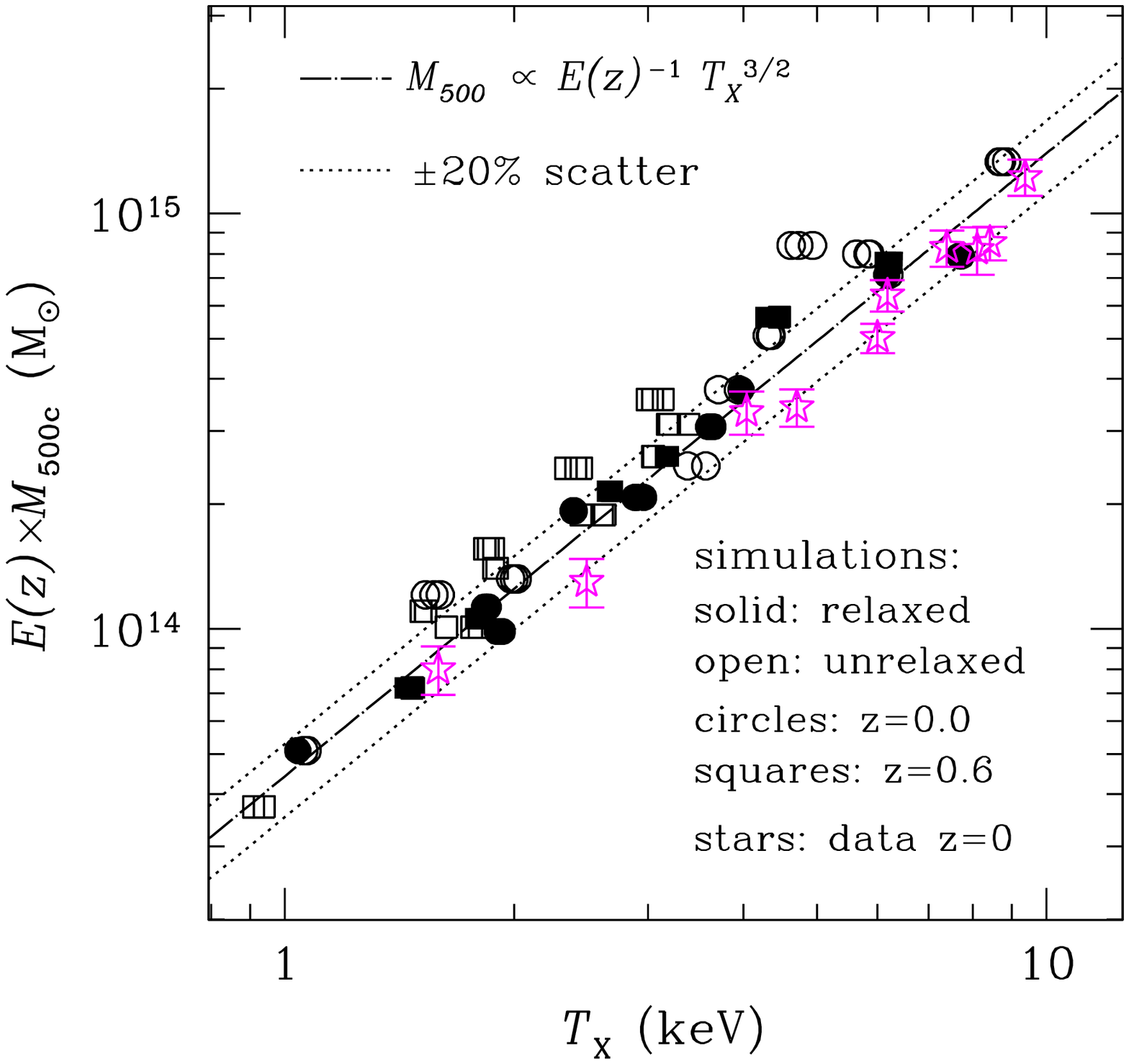} \hspace{-0.6cm}
  \includegraphics[height=6.1cm]{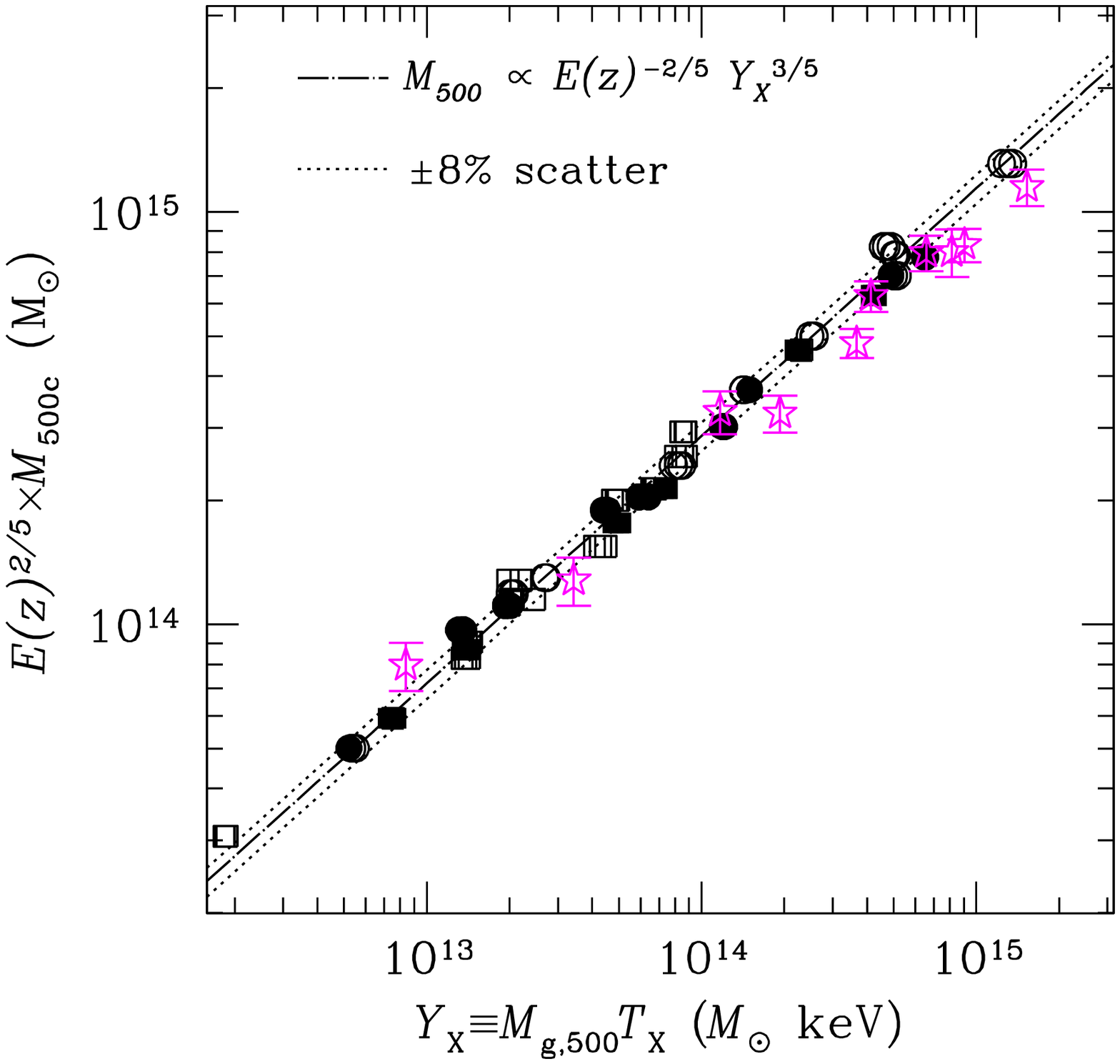}
\caption{Correlation between the total mass, $M_{500c}$ and X-ray spectral
  temperature, $T_X$ (\emph{left panel}) and the integrated X-ray pressure,
  $Y_X$ (\emph{right panel}). Separate symbols indicate relaxed and unrelaxed
  simulated clusters, and also z=0 and 0.6 samples.  The figures include
  points corresponding to three projections of each cluster.  The
  \emph{dot-dashed} lines are the power law relation with the self-similar
  slope fit for the sample of relaxed clusters.  The \emph{dotted} lines
  indicate 20\% and 8\% scatter, respectively. The data points with errorbars
  are Chandra measurements of nearby relaxed clusters.}
\label{fig4}     
\end{figure}

For cosmological application, it is important to understand the relations
between X-ray observables and cluster mass.  In Fig.~\ref{fig4}, we present
recent comparisons of two X-ray proxies for the cluster mass --- the spectral
temperature, $T_X$, and the new proxy, $Y_X$, defined as a simple product of
$T_X$ and $M_g$ \cite{kravtsov_etal06}.  Analogously to the integrated
Sunyaev-Zel'dovich flux, $Y_X$ is related to the total thermal energy of the
ICM.

The $M_{500}-T_X$ relation has a $\sim 20\%$ scatter in $M_{500}$ around the
mean relation, most of which is due to unrelaxed clusters.  The unrelaxed
clusters also have temperatures biased low for a given mass.  This is likely
because during mergers, the mass of the system has already increased but only
a fraction of the kinetic energy of merging systems is converted into the
thermal energy of gas, due to incomplete relaxation \cite{mathiesen_evrard01}.
The slope and evolution of the $M_{500}-T_X$ relation are also quite close to
the self-similar model.

The $M_{500}-Y_X$ relation shows the scatter of only $\approx 7\%$.  Note that
this value of scatter includes clusters at both low and high-redshifts and
both relaxed and unrelaxed systems. In fact, the scatter in $M_{500}-Y_X$ for
relaxed and unrelaxed systems is indistinguishable within the errors. $Y_X$ is
therefore a robust mass indicator with remarkably low scatter in $M_{500}$ for
fixed $Y_X$, regardless of whether the clusters are relaxed or not.  The
redshift evolution of the $Y_X-M_{500}$ relation is also close to the simple
self-similar prediction, which makes this indicator a very attractive
observable for studies of cluster mass function with X-ray selected samples,
because it indicates that the redshift evolution can be parameterized using a
simple, well-motivated function.

Finally, the results of the simulations are compared to the
observational results.  In both relations, the observed clusters show
a tight correlation with a slope close to the self-similar value.  The
normalization for our simulated sample agrees with the \emph{Chandra}
measurements to $\approx 10-15\%$. This is a considerable improvement
given that significant disagreement existed just several years ago
\cite{finoguenov_etal01,pierpaoli_etal03}. The residual systematic
difference in the normalization is likely caused by non-thermal
pressure support from bulk gas motions, which is unaccounted for in
X-ray hydrostatic mass estimates. The much improved agreement of
simulations and observations in these relations gives us confidence
that the clusters formed in modern simulations are sufficiently
realistic and thus can be meaningfully used for interpretation of
observations. The existence of tight relations of X-ray observables,
such as $Y_X$, and total cluster mass and the simple redshift
evolution of these relations hold promise for the use of clusters as
cosmological probes.

\begin{thebibliography}{99.}
  
\bibitem{vikhlinin_etal05c} {Vikhlinin}, A., {Markevitch}, M., {Murray},
  S.~S., {Jones}, C., {Forman}, W., \& {Van Speybroeck}, L. 2005, ApJ, 628,
  655

\bibitem{vikhlinin_etal06} {Vikhlinin}, A., {Kravtsov}, A., {Forman}, W.,
  {Jones}, C., {Markevitch}, M., {Murray}, S., \& {Van~Speybroeck}, L. 2006,
  ApJ, 640, 691

\bibitem{nagai_etal06} {Nagai}, D., {Kravtsov}, A.V., \& {Vikhlinin}, A.,
  2006, ApJ, in press (astro-ph/0609247)

\bibitem{faltenbacher_etal05} {Faltenbacher}, A., {Kravtsov}, A.~V., {Nagai},
  D., \& {Gottl{\"o}ber}, S.  2005, MNRAS, 358, 139
 
\bibitem{rasia_etal06} {Rasia}, E., {Ettori}, S., {Moscardini}, L.,
  {Mazzotta}, P., {Borgani}, S., {Dolag}, K., {Tormen}, G., {Cheng}, L.~M., \&
  {Diaferio}, A. 2006, MNRAS, 369, 2013
  
\bibitem{lau_etal06} {Lau}, E., {Kravtsov}, A.~V., \& {Nagai}, D. 2006, in
  preparation

\bibitem{mazzotta_etal04} {Mazzotta}, P., {Rasia}, E., {Moscardini}, L., \&
  {Tormen}, G. 2004, MNRAS, 354, 10
  
\bibitem{vikhlinin_etal06b} {Vikhlinin}, A. 2006, ApJ, 640, 710

\bibitem{voit_etal01} {Voit}, G.~M., {Bryan}, G.~L. 2001, Nature, 2001, 414,
  425
  
\bibitem{kravtsov_etal06} {Kravtsov}, A.~V., {Vikhlinin}, A.~A., \& {Nagai},
  D. 2006, ApJ, 650, 128

\bibitem{mathiesen_evrard01} {Mathiesen}, B.~F. \& {Evrard}, A.~E. 2001, ApJ,
  546, 100

\bibitem{finoguenov_etal01} {Finoguenov}, A., {Reiprich}, T.~H., \&
  {B{\"o}hringer}, H. 2001, A\&A, 368, 749
  
\bibitem{pierpaoli_etal03} {Pierpaoli}, E., {Borgani}, S., {Scott}, D., \&
  {White}, M. 2003, MNRAS, 342, 163

\end{thebibliography}
%


\printindex
\end{document}